\documentclass[showpacs,preprintnumbers]{revtex4}
\usepackage{amssymb}
\usepackage{amsmath}
\usepackage{graphicx}
\usepackage{dcolumn}
\usepackage{bm}
\usepackage{color}

\begin{document}

\title{Precisely position- and angular-controllable optical trapping and
manipulation via a single vortex-pair beam }
\author{Jisen Wen}
\thanks{J. S. Wen and B. J. Gao contributed equally to this work.}
\author{Binjie Gao}
\thanks{J. S. Wen and B. J. Gao contributed equally to this work.}
\author{Dadong Liu}
\author{Guiyuan Zhu}
\author{Li-Gang Wang}
\affiliation{Department of Physics, Zhejiang University, Hangzhou 310027, China}
\email{sxwlg@yahoo.com}

\begin{abstract}
Optical trapping and manipulation using structured laser beams now attract increasing attention in many
areas including biology, atomic science, and nanofabrication. Here we
propose and experimentally demonstrate the use of a single vortex-pair beam in
optical trapping and manipulation. Using the focal properties of such vortex-pair beams, we successfully manipulate two spherical
microparticles simultaneously, and obtain the precise
position-control on the microparticles by adjusting the off-axis parameter $a$ of the vortex-pair beam in its
initial phase plane. Furthermore, the high-precision angular-controllable rotation of cylindrical microrods is also achieved at will by rotating the initial phase structure
of such vortex-pair beams, like an optical wrench due to two focused bright spots at the focal plane of objective lens. Our result
provides an alternative manipulation of microparticles and may have potential applications in biological area, and optically driven micromachines or motors
\end{abstract}

\date{\today }
\pacs{42.65.Jx, 41.85.-p, 42.60.Jf, 42.25.-p, 42.25.Bs, 42.40.Eq, 42.62.-b}
\maketitle




\section{Introduction}

Optical trapping and manipulation have been widely used in a variety of
areas including atom cooling \cite{Chu1986,Kuga1997}, molecular biology \cite%
{Ashkin1987}, nanotechnology \cite{Jeffries2007,Marago2013}, as well as
other disciplines. Ashkin pioneered the investigation on optical force
produced by a laser beam
in 1970
\cite{Ashkin1970}. In 1986, Ashkin \emph{et al}. realized optical trapping of
dielectric particles by the gradient force from a single beam, which is
named as optical tweezers \cite{Ashkin1986}. Since then, optical tweezers become a
powerful and flexible tool for trapping and manipulating the
micrometre-sized objects. More importantly,
optical tweezers have many new progress, such as holographic
optical tweezers (HOTs) \cite{Dufresneg1998,Dufresneg2001,Curtis2002,Grier2003}, three-dimensional trapping of microparticles \cite{Rodrigo2015}, transportation of the micro-objects in the air \cite{Hadad2018}, and surface plasmon-based nano-optical tweezers \cite{Wang2018}.
The usual method of trapping microparticles often uses the gradient force of
a focused Gaussian beam. With the invention of new kinds of both scalar and
vectorial optical beams, many structured light beams are applied to optical
trapping, like Laguerre-Gaussian beams \cite{Neil2002-1,Jennifer2003,Chai2012}%
, Bessel beams \cite{Arlt2001,McGloin2003}, partially coherent light beams
\cite{Wang2007-2}, Airy beams \cite{Zhang2011}, helico-conical beams \cite%
{Daria2011}, radially polarized beams \cite{Yan2007}, cylindrical vector
beams \cite{Kozawa2010}, and Poincar\'{e} beams \cite{Wang2012}. In addition,
not only structured beams but also pulsed light can be used to trap linear
microparticles \cite{Ambardekar2005,Wang2007-1,Wang2011} and nonlinear
microparticles \cite{Devi2011,Gong2018}.

In many applications, it is required to manipulate the position of trapped microparticles or rotate them. In 2001, Dufresne \emph{et al.} designed the arrays of HOTs for trapping hundreds of
particles simultaneously and the position of trap is controllable
\cite{Dufresneg2001}. This method is also applied to manipulate the position
of neutral atoms \cite{Bergamini2004}. Leonardo \emph{et al.} also proposed
an iterative algorithm to make holograms for generating optical traps for
arbitrary three-dimensional trapping \cite{Leonardo2007}. Meanwhile, vortex
beams always rotate microparticles due to their orbital angular momentum,
however, this rotation cannot be well suppressed and well controlled
\cite{Simpson1997}. Additionally, rotating the spiral interference pattern
by an interferometer of high accuracy \cite{Paterson2001} or a rectangular
aperture inserted in the beam axis \cite{Neil2002-2} can also rotate
microparticles. Moreover, Zhang \emph{et al.} have theoretically proposed
optical doughnuts for rotating microparticles by rotating a modified vortex
phase \cite{Zhang2003}.

On the other hand, as the shape complexity of microparticles increases, like cylindrical microrods, trapping and
rotating these microparticles become more difficult. Cylindrical microrods
have been used in practical applications such as scanning optical probes in
force microscope \cite{Nakayama2007} and the fabrication of nanoelectronic
devices \cite{Yu2004}. Consequently, there has been also increasing interest
in the trapping of cylindrical microrods. For instance, Gauthier \emph{et al.} have experimentally observed that cylinders can be
manipulated and rotated in the transverse plane about the optical axis \cite{Gauthier1999}.
Other approaches including using a dual-beam fiber trap with one non-rotationally symmetric trapping
beam \cite{Kreysing2008}, a line optical trap \cite{Lee2009}, and shape-induced optical forces \cite{Phillips2014} have been used to manipulate micro-sized objects like cylinders or nanowires. Although HOTs exhibit great advantages for precisely rotating the cylinders \cite{Agarwal2005,Gao2019}, the novel methods for effectively trapping of the non-spherical particles are still expected in new practical applications.

The vortex-pair beam, a kind of structured light fields, contains a pair of
vortices at the initial phase plane, and it
was earlier studied analytically on its propagation in free space  by Guy in
1993 \cite{Guy1993}. Sometimes, vortex-pair beams, containing
opposite-charged vortices, are also known as vortex-dipole beams. Later, the annihilation phenomenon of vortex dipoles
has been discussed by Chen \emph{et al.} \cite{Chen2008}. The vortex-dipole beams propagating in various optical systems, for example, an astigmatic lens \cite{Yan2009,Reddy2014}, a high numerical-aperture system \cite{Chen2011,Zhao2018} and half-plane \cite{He2011,Singh2013}, have also been investigated.
It is also worth mentioning that the vortex-pair beam of electrons has been
experimentally realized by nanofabricated hologram grating in 2013 \cite%
{Hasegawa2013}. Although these works related to the evolution
properties of the vortex-pair beams have been investigated, the application of
the vortex-pair beams in optical trapping has never been realized yet.

In this work, we have experimentally realized the stable tapping of
spherical microparticles and cylindrical microrods based on a single
vortex-pair beam with two positive topological charges. It shows that the
precise control of spherical microparticles with their positions in one
direction can be realized. We have also demonstrated the controllable
rotation of cylindrical microrods by externally rotating the vortex-pair beam. To the best of our knowledge, it is
the first time to adopt a vortex-pair beam to realize the controllable
trapping and manipulation of the micrometre-sized objects. This method could
provide a high-precision and reliable control on the trapping of
microparticles and is potentially useful for many other areas.

\section{Theoretical description}

First, let us introduce the description of such vortex-pair beam. In
experiments, one can pass a Gaussian beam to be incident on a designed phase
plate with a pair of vortices. The light field at the initial plane can be
expressed as,
\begin{equation}
E_{\text{i}}(u,v)=G_{0}\exp (-\frac{u^{2}+v^{2}}{w_{0}^{2}})\phi (u,v),
\label{E_input)}
\end{equation}%
where $G_{0}$ is a constant related to the input power of light, $w_{0}$ is the beam
width of the incident Gaussian beam, $(u,v)$ are the Cartesian coordinates at the initial plane and $\phi (u,v)$ is the
initial pure phase function of the vortex-pair phase given by
\begin{equation}
\phi (u,v)=[\frac{u-a+iv}{\sqrt{(u-a)^{2}+v^{2}}}]^{m_{1}}[\frac{u+a+iv}{%
\sqrt{(u+a)^{2}+v^{2}}}]^{m_{2}},  \label{phase}
\end{equation}%
where $m_{1,2}$ are the integer topological charges of two vortices, and $a$
is the initial off-axis distance of each vortex. Note that the vortex-pair
phase is not rotationally symmetric, which is different from a single
integer vortex beam. When light propagates in a linear optical system, under
the paraxial approximation, the output field can be calculated by the
following Collins formula \cite{Collins1970,Wang2000}
\begin{align}
E_{\text{o}}(x,y,z)& \!=\!\frac{\exp (ikL)}{i\lambda B}\overset{+\infty }{%
\underset{-\infty }{\iint }}E_{\text{i}}(u,v)\exp \{\frac{ik}{2B}  \notag
\\
& \!\times \![A(u^{2}\!+\!v^{2})\!+\!D(x^{2}+y^{2})\!-\!2(ux\!+\!vy)]\}dudv,
\label{E_out}
\end{align}%
where $A$, $B$, and $D$ are the elements of a $2\times 2$ ray transfer
matrix $\left(
\begin{smallmatrix}
A & B \\
C & D%
\end{smallmatrix}%
\right) $ describing the total linear optical system from the input to
output planes, $L$ is the eikonal along the propagation axis, and $k=2\pi/\lambda$ is the wave number of
light. It is hard to obtain the analytical result, thus we have numerically solved
Eq. (\ref{E_out}) by using the trapezoidal numerical integration, which can be directly performed
on MATLAB software. In order to make sure the output fields correct, the discretization of the integrand function (its main part is the initial field expressed by Eq. (\ref{E_out})) has the small integral intervals, taking 1000 sampling data within the range of (-5$w_{0}$, 5$w_{0}$)
in the $x$ and $y$
directions, respectively, and there are total $10^{6}$ sampling points in the initial
plane. The requirement of the discretization procedure is to obtain the
stable solution of the output field with high precision, and the relative
error in our calculation is controlled to be smaller than $10^{-6}$, which
is sufficient to present correctly the field evolution in the focused
optical system.

\section{Results and discussions}

\begin{figure}[htb]
\centering
\includegraphics[width=13cm]{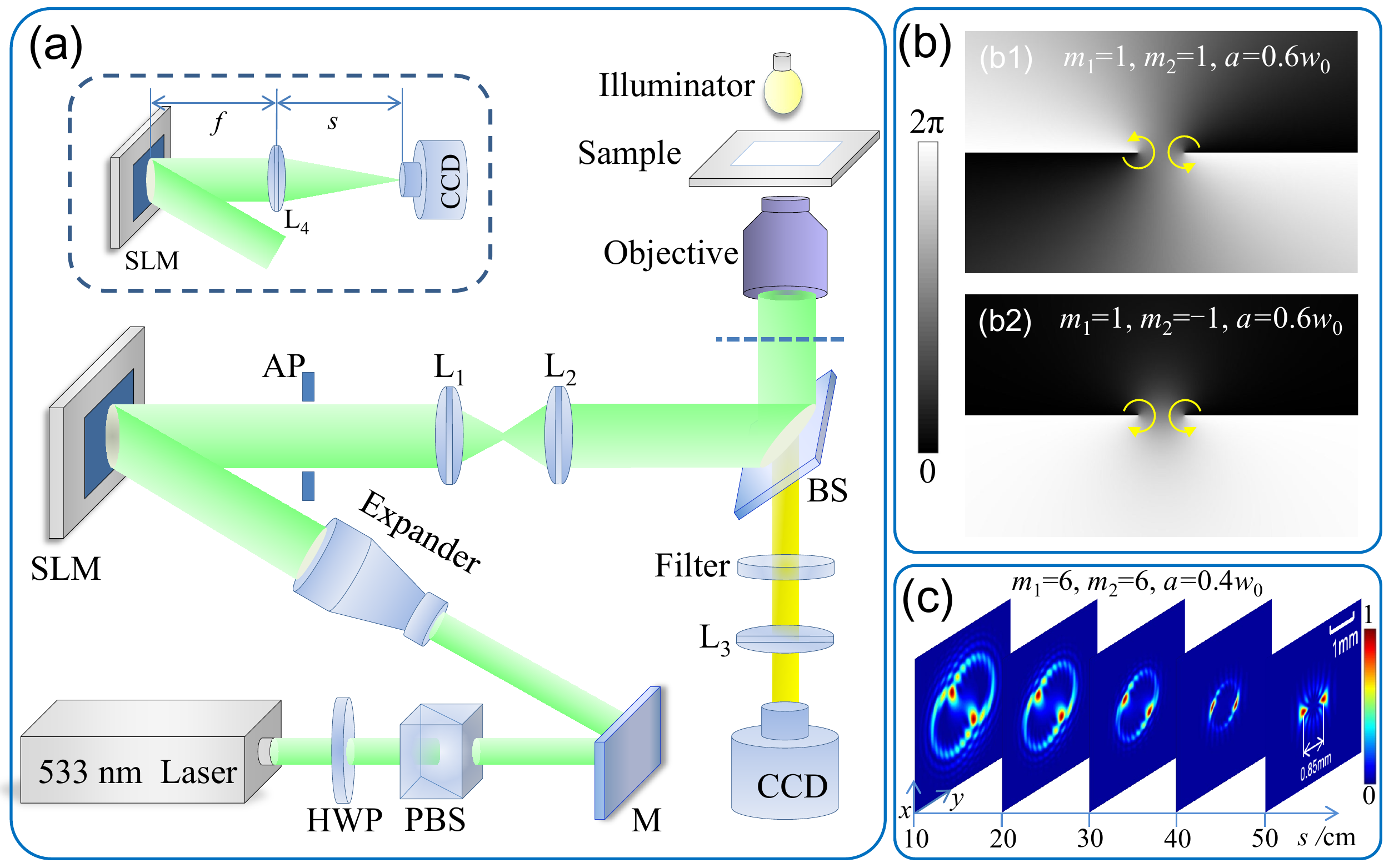}
\caption{(a) Schematic of optical trapping system. Notations are: HWP,
half-wave plate; PBS, polarized beam splitter; SLM, spatial light modulator;
AP, Aperture; BS, beam splitter; M, mirror; L, lens.
The dash line below the objective lens represents the position where we have measured the power of the incident
beam. The left-top inset in (a) shows the propagation of the vortex-pair beams in a 2-$f$
lens system with the focal length $f=50$ cm, which is used to explore the
focal property of such beams by changing $s$. (b) Two examples of vortex-pair phase with
different topological charges (b1, top) $m_{1}$=1, $m_{2}$=1 and (b2, bottom) $m_{1}$=1, $%
m_{2}$=-1. (c) Numerical results for the evolution of the vortex-pair beam
at different $s$ under the parameters $m_{1}$=6, $m_{2}$=6, $a$=0.4$w_{0}$ in a 2-$f$
system as shown in the inset of (a). Note that in our experiment the laser is expended to have $w_{0}=1.5$ mm before the SLM,
and $s$ in the inset of (a) is the distance from the lens to the CCD.}
\label{Fig1}
\end{figure}
We use this kind of beams to trap and manipulate microparticles, and the
experimental setup is shown in Fig. 1(a). The initial Gaussian beam with the
wavelength $\lambda =533$ nm is incident on a phase-only spatial light
modulator (SLM, Holoeye Pluto-2-VIS-056) with high resolution (1920$\times $%
1080 pixel) and 8 $\mu $m pixel pitch, which acts as a phase diffractive optical element.
The incident Gaussian
beam is expended by the expander, and the beam width $w_{0}$ is enlarged to be $w_{0}=1.5$ mm which is measured by a camera-based laser beam profiler (Ophir, SP928 Beam Profiling Camera). The beam waist $w_{0}=1.5$ mm is also used in numerical calculation. The polarized beam splitter combined with the half-wave plate is to realize the horizontal polarization and the
controllable power of the laser beam. Here the phase pattern on the SLM needs to be corrected due to the imperfect flatness of SLM's surface and the wavefront of incident beam. We have set the aberration correction function to generate the background compensation image by using the SLM pattern generator software at the beginning of experiment. We would also like to emphasize that in our experiment, the blazed grating phase (also known as prism phase) is applied on the SLM which can also be directly generated by the Holoeye SLM pattern generator software. The parameters of the blazed grating phase can be set by the software and the external designed vortex-pair phase distribution like Fig. 1(b) is loaded independently. The more detail operations can be found on the manual book of the SLM \cite{SLM056}.
Two examples of the vortex-pair phase with topological charges $m_{1}=m_{2}=1
$ and $m_{1}=-m_{2}=1$ are shown in Fig. \ref{Fig1}(b). The modulated first-order diffraction beam, which
is the vortex-pair beam, is selected by an aperture. The vortex-pair beam
then goes through a objective
lens($\times$100, NA=1.25; or $\times$40, NA=0.6) and is focused on the sample which is placed on a three-dimensional adjustable platform. The trapping system is configured with the vortex-pair beam directed upwards, which
allows easier access to realize adjustable platform. The sample is
illuminated by a white-light source from the top and is imaged onto a CCD camera. The filter
is used to block the green light reflected or scattered back from the cover
slip which has high intensity during the stable trapping process.

\begin{figure}[hbtp]
\centering
\includegraphics[width=10cm]{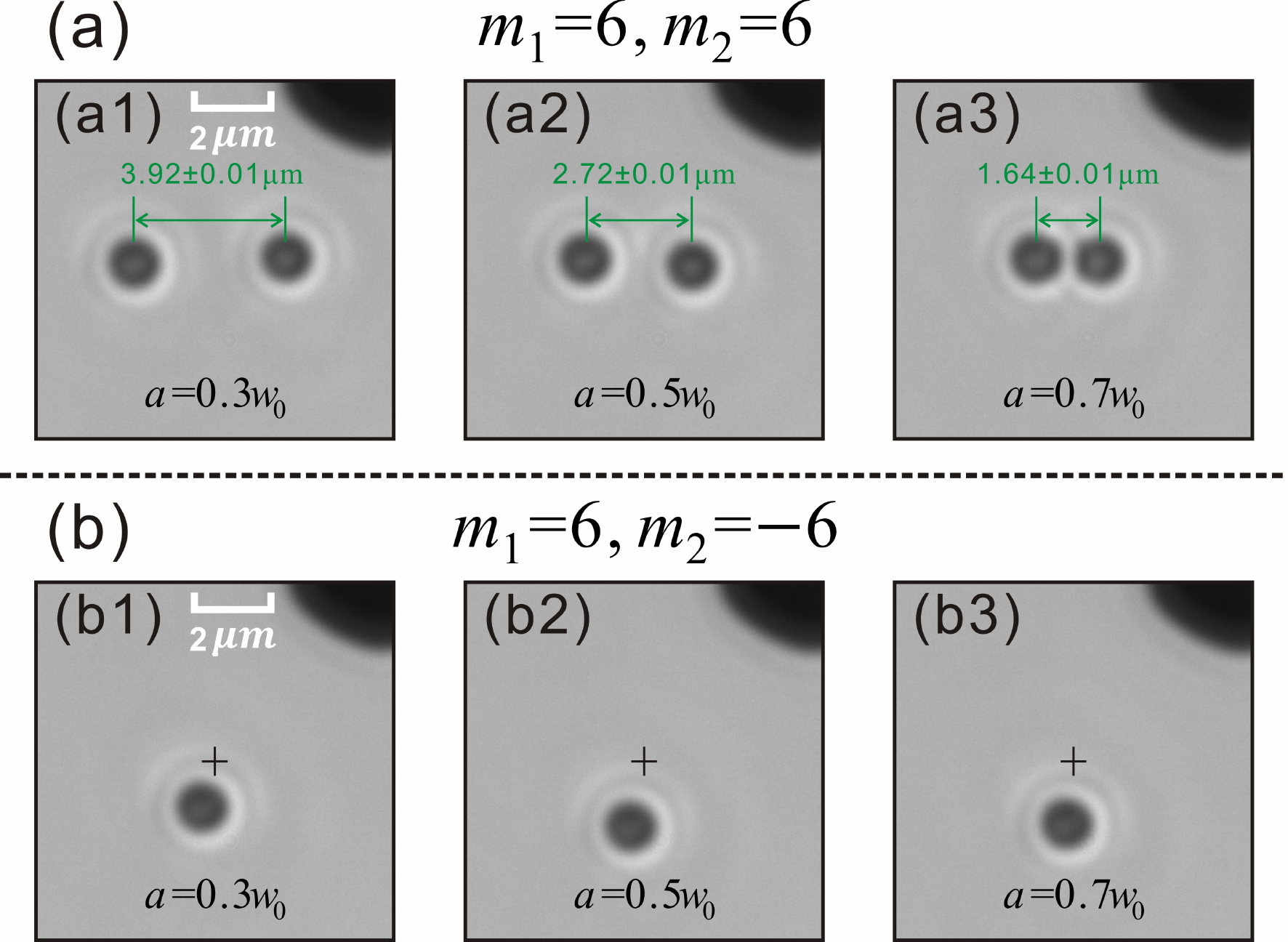}
\caption{
Camera snapshots of spherical microparticles inside the optical trap induced by the focused vortex-pair
beam with different topological charges (a) $m_{1}=6, m_{2}=6$ and (b) $m_{1}=6,m_{2}=-6$, through adjusting the off-axis parameter (a1, b1) $a=0.3w_{0}$, (a2,
b2) $a=0.5w_{0}$, (a3, b3) $a=0.7w_{0}$ in their initial phase structures.
Here $w_{0}=1.5$ mm, and the diameter of polystyrene microparticles $d_{p}=1 \mu m$. Note that the shadow on the right-top corner is one of the scale marks on the glass slide,
and the cross symbol in (b) is just for the convenient observation on the position of the single microparticle.}
\label{Fig2}
\end{figure}
Figure \ref{Fig2} shows the trapping and manipulation of spherical
microparticles using the vortex-pair beam. Here the polystyrene spheres (with diameter $d_{p}=1 \mu
$m \cite{tianjing}) are used as the probe samples in this
experiment. It is worth to mention that
the power of the incident vortex-pair beam to the objective lens is about 20 mW.
It is observed from Fig. \ref{Fig2}(a) that in the case of the vortex-pair
beam with positive topological charges $m_{1}=6$, $m_{2}=6$, two spherical
microparticles are trapped in the horizontal direction simultaneously. It
could be understood by Fig. 1(c) that when
the vortex-pair beam is focused by an objective lens, there exists two
bright spots at the focal plane, which can directly trap two microparticles.
When the parameter $a$ contained in the SLM phase structure is adjusted, the
position of the trapped spherical microparticles is also changed. For
instance, the separation distance $d_{s}$ of the two trapped spherical
microparticles is 3.92 $\mu $m with $a=0.3w_{0}$ as shown in Fig. \ref{Fig2}%
(a1), while it decreases to 1.64 $\mu $m under $a=0.7w_{0}$ displayed in Fig. %
\ref{Fig2}(a3). It seems that the larger $a$ is, the closer the separation $d_{s}$
of two trapped spherical microparticles is. Therefore, one can precisely control
the distance between two interested microparticles by this method. However,
in the case of $m_{1}=6$, $m_{2}=-6$, it is seen in Fig. \ref{Fig2}(b) that
only one spherical microparticle is trapped as compared with the case of $%
m_{1}=6$, $m_{2}=6$. Furthermore, adjusting $a$ makes no clear difference on
the trapping results, except for a slight change of the microparticle's position.

\begin{figure}[hbtp]
\centering
\includegraphics[width=14cm]{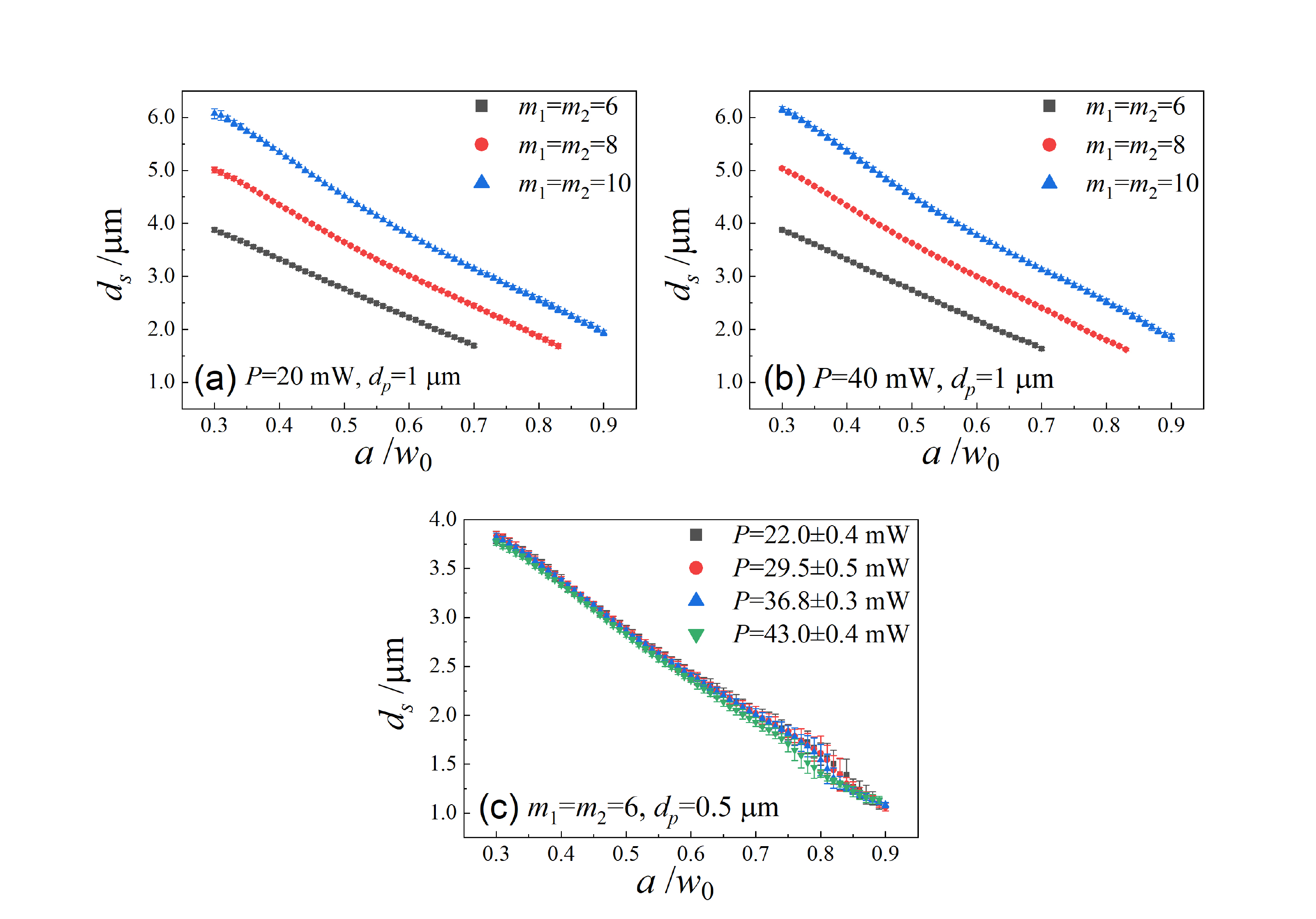}
\caption{Experimentally measured separation distance $d_{s}$ between two trapped
spherical microparticles as a function of the off-axis parameter $a$ for
different vortex-pair beams under both different incident power $P$ and different sizes of microparticles.
In (a, b)  the incident power $P$ is 20 mW and 40 mW, respectively, and the diameter of microparticles is fixed to be $d_{p}=1 \mu $m.
In (c), the values of $m_{1}$, $m_{2}$ are both equal to 6, $d_{p}=0.5 \mu $m, and the different laser power is shown in figure.}
\label{Fig3}
\end{figure}

Figure \ref{Fig3} further shows the dependence of the separation distance $d_{s}$
between two trapped spherical microparticles on the control parameter of the
off-axis distance $a$ in the initial vortex-pair phase.
Here, we used the polystyrene microspheres with diameter $d_{p}=1 \mu$m, $0.5 \mu$m \cite{tianjing}. When the vortex-pair
beam has two positive topological charges, for example, in the case of $%
m_{1}=m_{2}=6$, the vortex-pair beam can be focused into two bright spots as
shown in Fig. 1(c) (at $s=f=50$ cm). These two bright spots with high
intensity peaks lead to the strong transverse gradient forces for trapping microparticles.
Thus two spherical microparticles are trapped, as shown in Figs. 2(a1) to
2(a3). The separation distance $d_{s}$ between these two microparticles can be
manipulated precisely by changing the value of $a$. For large-size microparticles of the cases in Figs. 3(a) and 3(b),
whatever the power is, for examples when $P=20$ mW and $P=40$ mW (we have also measured the effect under other incident power),
the separation distance $d_{s}$ between the two trapped microparticles decreases linearly as $a$ increases,
and such linearity of $d_{s}$ vs $a$ keeps very well for those microparticles with $d_{p}=1 \mu$m, 2 $\mu$m, and 3 $\mu$m.
For the experimental results for $d_{p}=2 \mu$m, and 3 $\mu$m are not presented here since the results are similar.
However, in Fig. 3(c), we have also observed the clear deviation of $d_{s}$ from the linearity vs $a$ under different incident power
when we used the smaller particles with $d_{p}=0.5 \mu$m. From Fig. 3(c), as the incident power $P$ decreases, from 43 mW to 22 mW, under the same conditions of
other parameters, the deviation from the linearity of the value $d_{s}$ vs $a$ is more significant around the value of $a=0.8 w_{0}$, especially for $P=22$ mW.
The experiment measurement for such smaller sized microparticles cannot be performed under the lower power because
the microparticles will escape from the optical trap for the lower power.
We think that such deviation phenomenon is due to the influence of the optical binding effect between two microparticles \cite{Burns1989,Ng2005}.
All the experimental data are measured from the images like Figs. 2(a1) to 2(a3) in different cases of $%
m_{1}=m_{2}=6$, $m_{1}=m_{2}=8$, and $m_{1}=m_{2}=10$. The error bars are
determined from multiple repeated measurements and are very small in the linear range of $d_{s}$ vs $a$. From
Figs. 3(a) and 3(b), we can see that one can manipulate the separation between trapped
particles both precisely and linearly through changing the external
parameter contained in the initial phase of such vortex-pair beams,
and it is also better to choose a vortex-pair beam with two larger
topological charge numbers for realization of a large position-controllable
manipulation range between two trapped samples in practice.

For the smaller microparticles with $d_{p}=0.5 \mu $m, there may appear the optical binding effect. It is interesting
and worth to further investigate how the optical binding effect impact on the optical trapping and manipulation
for such structured inhomogenous focused light fields, since there maybe exist the competitive mechanism
between optical binding effect and the optical gradient forces.

\begin{figure}[hbt]
\centering
\includegraphics[width=14cm]{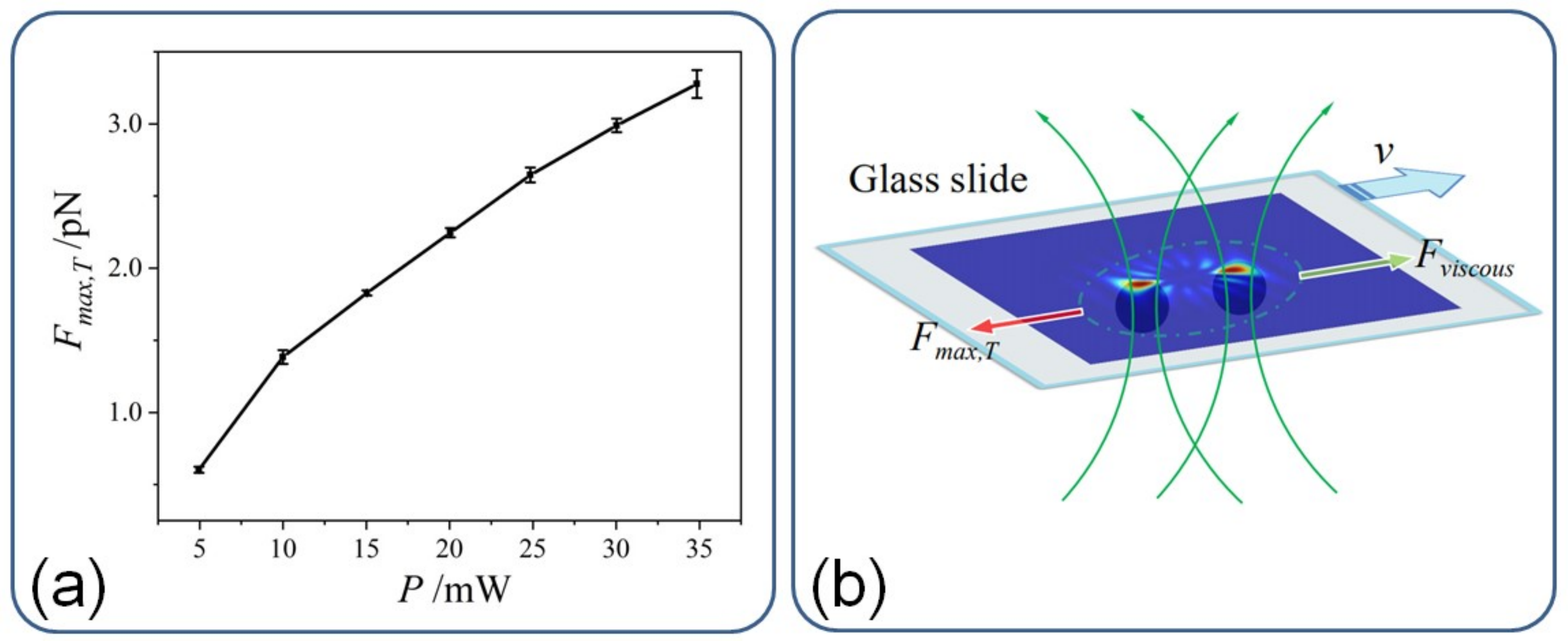}
\caption{(a) Experimental measurement of the maximal transverse optical trapping force of the vortex-pair beam with
the same positive topological charges $m_{1}=m_{2}=6$, the off-axis distance $a=0.4w_{0}$, and $w_{0}=1.5$
mm. (b) A schematic diagram to measure the maximal transverse optical force on the two microparticles.
Note that the glass slide is moving from the left to right, denoted by the arrow of $v$,
and in the horizontal direction there are mainly two opposite forces: the transverse optical force and the viscous force on microparticles.
Once the speed of the glass slide increases to reach a critical value, the trapped microparticles will escape from the optical trap due to
the increasing viscous force.}
\label{Fig4}
\end{figure}

In Fig. \ref{Fig4} we have experimentally measured the maximal magnitude of the transverse optical trapping force by
the method of fluid mechanics \cite{Stoke2005}. It shows that the magnitude of the measured maximal optical trapping force increases
as a function of the input laser power. When the laser power changes from 15mW to 35 mW, the optical trapping force is
roughly proportional to the power but not fully linear. Because the frictional force between the particles and the glass
surface may have certain influence on the measurement since the scattering force in the propagation direction can push
the particles close to the bottom surface of glass slide (note that our sample is located between the objective lens and glass slide).
As the power increases, the frictional force may also increase.

\begin{figure}[t!]
\centering
\includegraphics[width=14cm]{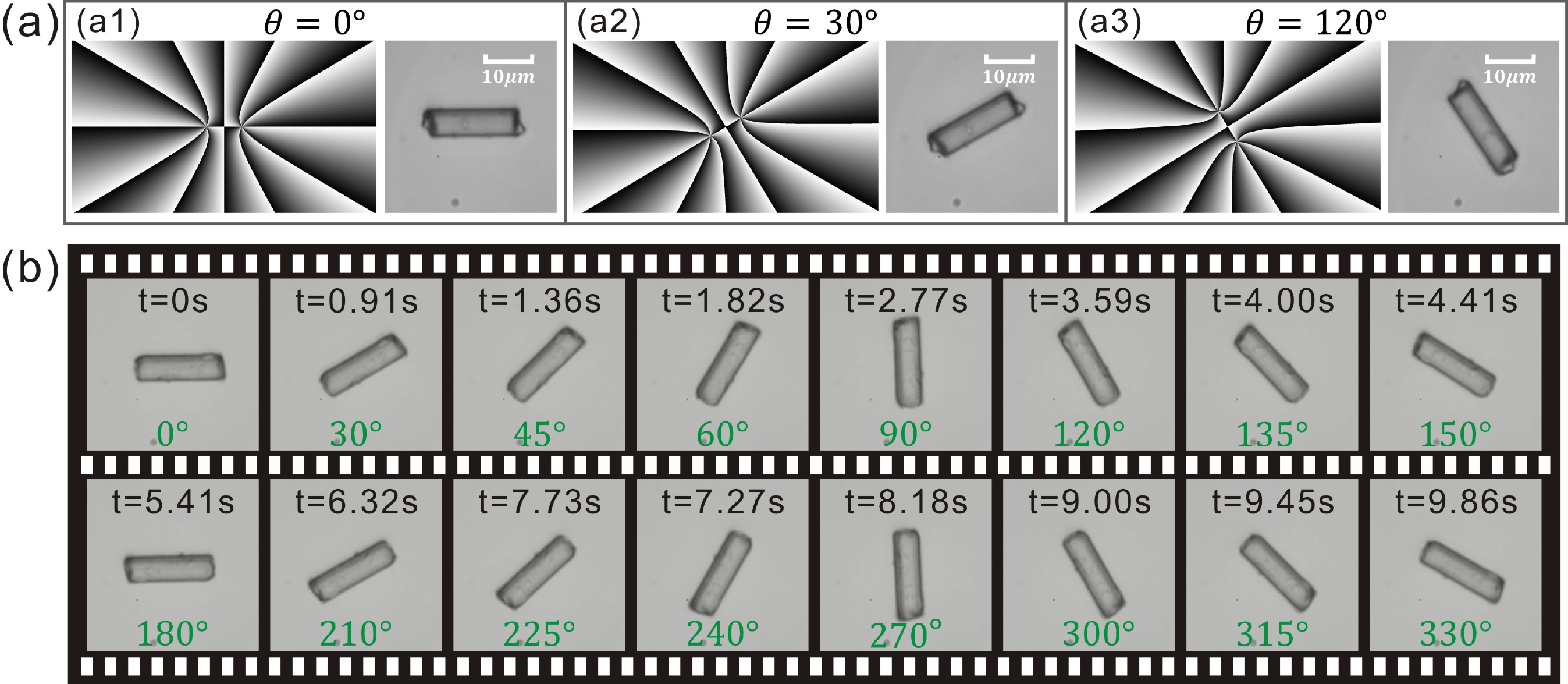}
\caption{Experimental realization of the angular-controllable rotation of
the cylindrical microrods. (a) Rotated initial phase distributions (left) of the
vortex-pair beams loaded on SLM and rotated cylindrical microrod (right)
trapped by the vortex-pair beam with the specific angle (a1) $\protect\theta %
=0^{\circ }$, (a2) $\protect\theta =30^{\circ }$, and (a3) $\protect\theta %
=120^{\circ }$. (b) Camera snapshots of continuously rotating cylindrical
microrod trapped by the vortex-pair beam with the same positive topological
charges $m_{1}=m_{2}=6$, the off-axis distance $a=0.6w_{0}$, and $w_{0}=1.5$
mm. The corresponding elapsing time $t$ and the rotating angle are denoted in each snapshot,
and there is a video that can be seen in Visualization 1. }
\label{Fig5}
\end{figure}

It is observed from Fig. \ref{Fig1}(b) that the initial phase structure of
the vortex-pair beam is rotationally asymmetric. Therefore, naturally, this
kind of beam becomes a perfect light source for controllable rotation of the
trapped microparticle by rotating the beam. Experimentally, we can rotate
the initial phase loaded on the SLM which is equivalent to rotate the
vortex-pair beam. Figure \ref{Fig5} gives the experimental results on the
controllable rotation of a cylindrical microrod realized by rotating the
initial phase. The rotated initial phase is realized by the transform of $\left\{
\begin{array}{c}
u^{\prime }=u\cos \theta (t)+v\sin \theta (t) \\
v^{\prime }=-u\sin \theta (t)+v\cos \theta (t)%
\end{array}%
\right.$, which is the rotation transformation of the rectangular coordinates.
The rotation angle $\theta (t)$ can be arbitrarily designed ranged from $0^{\circ }$ to $%
360^{\circ }$, and $t$ represents the elapsed time of encoding the initial
vortex-pair phase with different rotated angle on the SLM, and $\overline{%
\omega }=\Delta \theta (t)/\Delta t$ can be the average angular speed of
rotation for a vortex-pair beam, where $\Delta \theta $ and $\Delta t$
represent the step of the rotated angle of encoding phase image and
the encoding time step, respectively, e. g., 100 images of phase distributions are designed
with the rotated angle uniformly from $0^{\circ }$ to $360^{\circ }$, thus $%
\Delta \theta =3.6^{\circ }$, and the elapsed time for encoding 100 phase
images is set to 10 s so that it means $\Delta t=100$ ms. In this
experiment, a specially designed
cylindrical microrod with length 19 $\mu $m and diameter 4.5 $\mu $m made of
silica is used \cite{zibo} and  the power of the incident
vortex-pair beam to the objective lens now is 35 mW. It is observed from
Fig. 5(a2) that when the vortex-pair beam is rotated by $30^{\circ }
$, via rotating the initial phase on SLM, the trapped cylindrical microrod
is also rotated by near $30^{\circ }$ as compared with the non-rotation case
as seen in Fig. 5(a1). As expected, Fig. 5(a3) also shows
that the rotation angle of the trapped cylindrical microrod is consistent
well with the rotation angle of the initial vortex pair phase.
As compared with the rotation induced by a vortex beam, this method shows high controllability,
and our method also decreases the complexity of rotation with the simpler experimental setup
as compared with the method in Ref. \cite{Paterson2001}. Thus, it can be
concluded that controllable rotation of trapping cylindrical microrod is
well realized and the rotation process can be precisely controlled.
Interestingly, a continuous rotation of the trapped cylindrical microrod is
experimentally performed (see Visualization 1). In Fig. 5(b), a
tweezed cylindrical microrod can be seen to rotate between the frames which
is extracted from the video (see Visualization 1) at some typical angles.
Here the average angular speed is $\overline{\omega }\approx 36^{\circ }$%
/sec.

\begin{figure}[htbp]
\centering
\includegraphics[width=14cm]{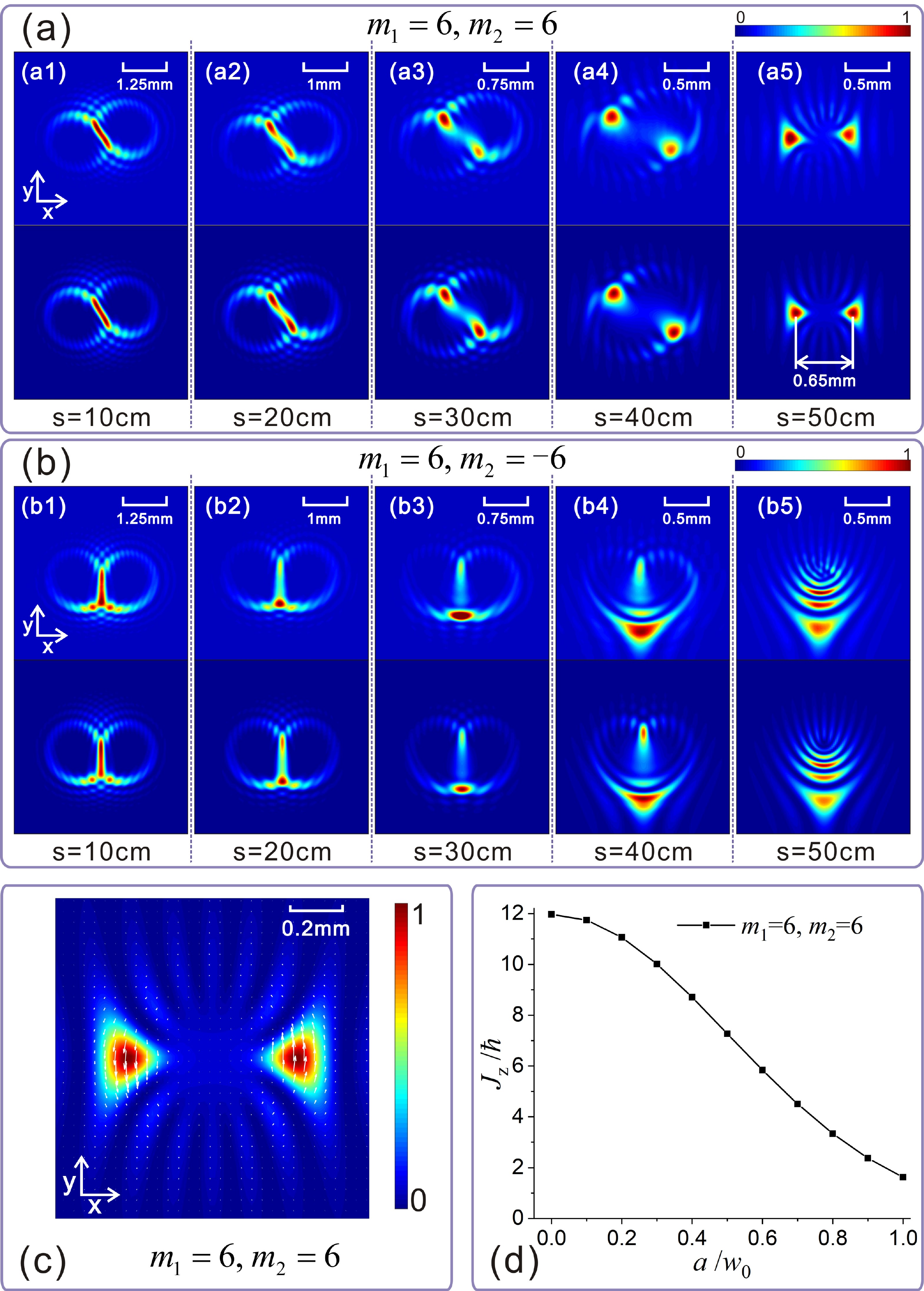}
\caption{The experimental (top) and numerical (bottom) results for the normalized intensity distributions
 of the focused vortex-pair beams with topological charges
(a) $m_{1}=m_{2}=6$ and (b) $m_{1}=6,m_{2}=-6$, at different propagation distances $s$ in a 2-$f$ lens
system under the off-axis distance $a=0.6w_{0}$, $w_{0}=1.5$ mm, and $f=50$ cm.
(c) The transverse energy flow of
vortex-pair beam with topological charge $m_{1}=m_{2}=6$.
(d) The orbital angular momentum of the vortex-pair beam with $m_{1}=m_{2}=6$ as a function of the off-axis parameter $a$.}
\label{Fig6}
\end{figure}
In order to explain the experimental results in Figs. 2,3, and 5, we have further measured
the intensity distributions of such vortex-pair beams in a focused system,
which can mimic the intensity distributions near the focal plane of the
objective lens in our optical tweezers systems. Figure 6 plots the measured
and numerical intensity distributions of the focused vortex-pair beams when
they gradually approach to their focal point at $s=f=50$ cm in the 2-$f$ lens system, which is shown in inset of Fig. 1(a). Here, we want
to emphasize that although there are some previous works related to the evolutions of the
vortex-pair beams \cite{Guy1993,Chen2011,He2011,Zhao2018}, strictly speaking,
those beams \cite{Guy1993,Chen2011,He2011,Zhao2018} with a pair of vortices
are different from our design since we have used a pure phase function given by
Eq. (\ref{phase}) and there is no amplitude modulation in our cases.
Surprisingly, little work has been done experimentally on investigating the
focusing property of such vortex-pair beams. Therefore, we have also performed the
experiment to systematically demonstrate the focal evolution of the
vortex-pair beam propagating in a 2-$f$ lens system, which is actually
equivalent to a objective lens system in the optical trapping system. The
transfer matrix of such a focal system for the inset of Fig. 1(a) is given by $\left(
\begin{smallmatrix}
A & B \\
C & D%
\end{smallmatrix}%
\right) =\left(
\begin{smallmatrix}
1-(s/f) & f \\
-1/f & 0%
\end{smallmatrix}%
\right) $. Substitute it into Eq. (\ref{E_out}), the output light field can
be obtained numerically. A lens with focal length $f=50$ cm is applied to
our experiment. Fig. 6 shows the intensity dynamics of the
vortex-pair beam with $a=0.6w_{0}$ at different propagation distances behind
the lens. In Fig. 6(a1), in the case of the vortex-pair beam with two
positive topological charges, there are two ring-like patterns resulted from
the diffraction by the off-axis vortices at the initial phase plane, and
these two ring-like patterns can interfere with each other and generate the
interference patterns. As $s$ closes to the focal
point, the strongest intensity peaks gradually appear near the overlapped
area where the original two ring-like patterns are overlapped. As the beam
propagates to focal plane, see Figs. 6(a3)-6(a4), both two strongest spots
rotate each other anticlockwise due to the role of the Gouy phase \cite%
{Guy1993}. Finally, the intensity distribution becomes two main bright spots
along the $x$ axis, which is symmetric about the $x$ and $y$ axes,
respectively. These two bright spots are very like two optical nails for
catching both microparticles and microrods firmly and precisely because of the strong transverse optical gradient forces, thus the
beam like an optical wrench can control the microrods precisely. When
 controlling the external parameters to rotate the beam generated by the SLM,
one can have the operations of positioning or rotating the targets as
discussed above. However, in the case of the vortex-pair beam with two
opposite topological charges, as shown in Figs. 6(b1) to 6(b5), there are
different behaviors for the intensity evolution due to the attractive
properties of these two opposite vortices, and the intensity distribution
near the focal plane simply becomes a single main bright spot with multiple
curved interference fringes. Such different focal behaviors lead to the
different trapping effect shown in Fig. 2. These two ring-like patterns
attract each other, thus the vortex-pair beam with opposite-sign charges
experiences destructive interference, and the intensity distribution at
focal plane is mainly located at the $y$ axis (which is symmetric about the $%
y$ axis but not $x$ axis). Although it can also rotate the cylindrical
microrod by rotating the beam, the rotation of the cylindrical microrod is
not as smooth and stable as the case of the beam with two positive
topological charges (see Visualization 2).
Fig. 6(c) shows the energy flow of the vortex-pair beam with topological
charge $m_{1}=m_{2}=6$ at focal plane. It is observed that energy flow at right is upward while the energy flow at left is downward.
Moreover, the total orbital angular momentum of the vortex-pair beam decreases as the off-axis distance $a$ increases as shown in Fig. 6(d)
which is similar to the case of a single off-axis vortex beam in Ref. \cite{Kotlyar2019}.
However, it should be emphasized that the rotation of the microrod here is realized by rotating the vortex-pair beam with two positive topological
charges based on the gradient force rather than the orbital angular momentum, since the orbital angular momentum here is weak and uncontrollable
to manipulate the rotation of the particles.
According to Ref. \cite{Wang2012,Harada1996}, we have also roughly calculated the transverse gradient force of the vortex-pair beam ($m_{1}=m_{2}=6,a=0.4w_{0}$) and the vortex beam (topological charge $\alpha =6$) under the parameter $w_0=1.5$  mm, $f=2$ mm and input power $P=20$ mW at the focal plane of a 2-$f$ system which indicates that the maximum gradient force of the vortex-pair beam is about 20$\%$ higher than the vortex beam.
\section{Conclusion}

In summary, we experimentally realized a kind of optical tweezers based on a
single vortex-pair beam with two positive charged numbers, which provides an
alternative candidate for the trapping and manipulation of microparticles
and microrods. It shows that two spherical-shaped microparticles are trapped
simultaneously using such kind of beams, and the separation distance between
the two trapped spherical shaped microparticles can be tuned linearly by the
off-axis distance of the vortex pair within the initial phase. Due to the
special properties of the focused vortex-pair beam, it offers a flexible
tool to simultaneously manipulate two trapped spherical microparticles and
high-precisely control the orientation of the cylindrical microrods, which
both can be adjusted by the external control parameters of such beams. The
approach may decrease the difficulty in locating or aligning multiple
microparticles and microrods since the design of vortex-pair phase is very
convenient. Our results may also have potential applications in many areas
like biological area and nanofabrication.


\begin{acknowledgments}
This work was supported by the National Natural Science Foundation of China
(NSFC) (grants No.11974309 and 11674284), National Key Research and
Development Program of China (No. 2017YFA0304202), Zhejiang Provincial
Natural Science Foundation of China under Grant No. LD18A040001, and the
Fundamental Research Funds for the Center Universities (No. 2019FZA3005).
\end{acknowledgments}

\end{document}